\documentclass[aps,prl,twocolumn,amsmath,amssymb,floatfix,bibnotes,preprintnumbers,superscriptaddress,longbibliography]{revtex4-2}
\usepackage[T1]{fontenc}
\usepackage[utf8]{inputenc}
\usepackage{amsmath}
\usepackage{graphicx,epstopdf}
\usepackage{gensymb}
\usepackage{float}
\usepackage{dcolumn,booktabs,microtype,afterpage}
\usepackage{amssymb}
\usepackage{sidecap}
\usepackage[mathlines]{lineno}
\makeatletter
\usepackage{color}
\usepackage[colorlinks,bookmarks=false,citecolor=darkblue,linkcolor=red,urlcolor=blue]{hyperref}

\definecolor{darkred}{rgb}{0.7,0.0,0.0}

\definecolor{darkblue}{rgb}{0,0.02,0.45}

\definecolor{darkgreen}{rgb}{0.02,0.45,0.0}

\definecolor{violet}{rgb}{0.8,0.2,0.6}

\DeclareUnicodeCharacter{2212}{-}

\begin{document}
	
	\title{Pressure-Induced Enhancement of Superfluid Density in Transition Metal Dichalcogenides with and without Charge Density Wave}
	
	\author{S.S.~Islam}
	\thanks{These authors contributed equally to the work.}
	\affiliation{PSI Center for Neutron and Muon Sciences CNM, 5232 Villigen PSI, Switzerland}
	
	\author{V.~Sazgari}
    \thanks{These authors contributed equally to the work.}
	\affiliation{PSI Center for Neutron and Muon Sciences CNM, 5232 Villigen PSI, Switzerland}
    
	\author{C.~Witteveen}
	\thanks{These authors contributed equally to the work.}
	\affiliation{Department of Quantum Matter Physics, University of Geneva, CH-1211 Geneva, Switzerland}
	\affiliation{Department of Physics, University of Zurich, Winterthurerstr. 190, 8057 Zurich, Switzerland}
		
	\author{J.N.~Graham}
	\affiliation{PSI Center for Neutron and Muon Sciences CNM, 5232 Villigen PSI, Switzerland}
		
	\author{O.~Gerguri}
	\affiliation{PSI Center for Neutron and Muon Sciences CNM, 5232 Villigen PSI, Switzerland}
	\affiliation{Department of Physics, University of Zurich, Winterthurerstr. 190, 8057 Zurich, Switzerland}

    \author{P. Kr\'{a}l}
	\affiliation{PSI Center for Neutron and Muon Sciences CNM, 5232 Villigen PSI, Switzerland}

    \author{M.~Bartkowiak}
	\affiliation{PSI Center for Neutron and Muon Sciences CNM, 5232 Villigen PSI, Switzerland}
        
	\author{H.~Luetkens}
	\affiliation{PSI Center for Neutron and Muon Sciences CNM, 5232 Villigen PSI, Switzerland}
	
	\author{R.~Khasanov}
	\affiliation{PSI Center for Neutron and Muon Sciences CNM, 5232 Villigen PSI, Switzerland}
	
	\author{F.O.~von~Rohr}
	\email{fabian.vonrohr@unige.ch}
	\affiliation{Department of Quantum Matter Physics, University of Geneva, CH-1211 Geneva, Switzerland}
	
	\author{Z.~Guguchia}
	\email{zurab.guguchia@psi.ch}
	\affiliation{PSI Center for Neutron and Muon Sciences CNM, 5232 Villigen PSI, Switzerland}
	
	\date{\today}
	
	\begin{abstract}
		
		Gaining a deeper understanding of the interplay between charge density wave (CDW) order and superconductivity in transition metal dichalcogenides (TMDs), particularly within the (4H/2H)-NbX$_{2}$ (X=Se,S) family, remains an open and intriguing challenge. A systematic microscopic study across various compounds in this family is therefore required to unravel this complex interplay. Here, we report on muon spin rotation and magnetotransport experiments investigating the effects of hydrostatic pressure on the superconducting transition temperature ($T_{\rm c}$), the temperature-dependent magnetic penetration depth ($\lambda_\mathrm{eff}$), and the charge density wave order (CDW) in two layered chalcogenide superconductors: 4H-NbSe$_{2}$, which exhibits CDW order, and 2H-NbS$_{2}$, which lacks such order. Our observations reveal a substantial 75$\%$ enhancement of the superfluid density ($n_{s}/m^{*}$) in 4H-NbSe$_{2}$ upon maximum applied pressure of $\sim2$\,GPa, surpassing that of 2H-NbSe$_{2}$. Despite the absence of CDW order, a sizeable 20$\%$ growth in superfluid density is also observed for 2H-NbS$_{2}$ under an applied pressure of 1.8\,GPa. Notably, the evaluated superconducting gaps in all these TMDs remain largely unaffected by changes in applied pressure, irrespective of pressure-induced partial suppression of CDW order in (4H/2H)-NbSe$_{2}$ or its general absence in 2H-NbS$_{2}$. These results underscore the complex nature of pressure-induced behaviors in these TMDs, challenging a simplistic view of competition solely between CDW order and superconductivity. Remarkably, the relationship between $n_{s}/m^{*}$ and $T_{\rm c}$ exhibits an unconventional correlation, indicating a noteworthy similarity with the behavior observed in cuprate, kagome, and iron-based superconductors.
			
	\end{abstract}
	\maketitle
	
	Transition metal dichalcogenides (TMDs)~\cite{vonRohr8465,Guguchia1082} represent a fascinating class of materials that have garnered significant attention in condensed matter physics due to their diverse electronic properties~\cite{Wang6960, Guguchia3672}. The superconducting state in TMDs has revealed an intriguing non-BCS behavior and unconventional superconducting states, both in bulk and monolayer form~\cite{Sipos960,Ribakeaax9480,Xi139,Zhang1425}. In particular, a correlation between the superfluid density and transition temperature ($T_{\rm c}$) serves as a notable manifestation of unconventional superconductivity in bulk 2H-NbSe$_2$~\cite{vonRohr8465}. Additionally, the reported upper critical field ($H_{c2}$) values for 4H-NbSe$_{2}$ [$\mu_{0}H_{c2}(0) = 26.5$\,T]~\cite{Zhou224518} and 2H-NbS$_{2}$ [$\mu_{0}H_{c2}(0)\|_{\rm ab} = 23$\,T]~\cite{Cho3676} are much higher than the Pauli paramagnetic limit [$\mu_{0}H_{\rm p}=1.84\times T_{\rm c}$\,(T/K)] expected for a weakly coupled BCS superconductor. 
	Recent advancements in the bulk TMD research have uncovered interesting phenomena, such as the observation of a Fulde-Ferrell-Larkin-Ovchinnikov (FFLO) state for magnetic fields applied exactly parallel to the $ab$-plane~\cite{Cho3676,Wan46}. The Ising-type superconductivity of monolayer 2H-NbSe$_2$ is, among other exceptional electronic properties, of particular interest~\cite{Xi139}. These discoveries challenge conventional paradigms and open up new avenues for exploring unconventional superconductivity in TMDs, thereby expanding our understanding of the complex interplay between various electronic phases.
	
	Furthermore, TMDs like 2H-NbSe$_2$~\cite{Yoshizawa056401} and 4H-NbSe$_2$~\cite{Zhou224518} exhibit CDW order in addition to superconductivity. The coexistence and interaction of these two phenomena raise intriguing questions about their origin, stability, and mutual influence. Pressure-induced phase transitions, which can suppress CDW order and alter superconducting behavior, offer valuable insights into the underlying mechanisms governing these phenomena~\cite{Berthier1393, Suderow117006, Feng7224,Moulding043392,Cao245125}. Therefore, understanding the response of TMDs under pressure provides crucial information about the delicate balance between the competing electronic phases and sheds light on the unconventional nature of their superconducting state.
	
	\begin{figure*}[t!]
		\centering
		\includegraphics[width=\linewidth]{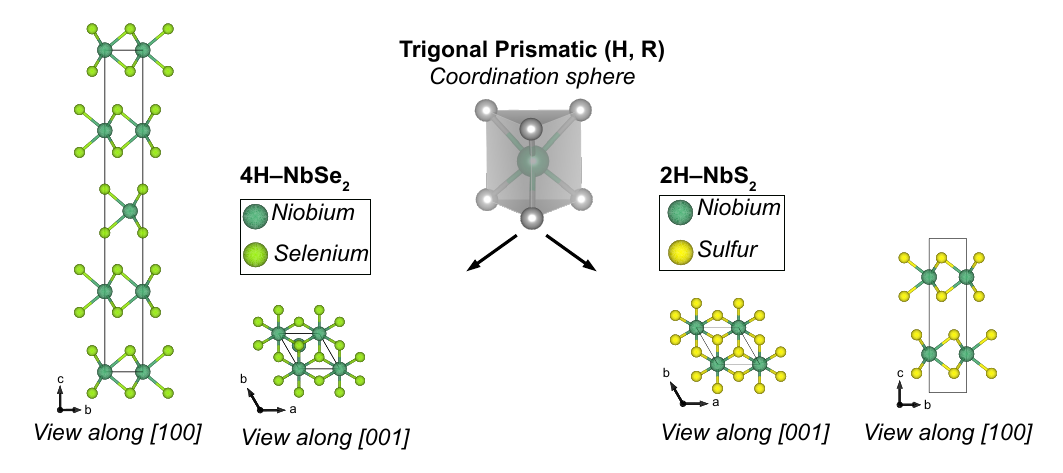}
		\vspace{-0.7cm}
		\caption{\textbf{Crystal structure.}
		Middle: Illustration of the trigonal prismatic coordination sphere of the chalcogen atoms (yellow: sulfur, light green: selenium) around the niobium (dark green). Left: Side and top view of the 4H-NbSe$_{2}$, with four layers per unit cell. Right: Top and side view of the 2H-NbS$_{2}$, with two layers per unit cell~\cite{Momma1272}.}
		\label{Fig1}
	\end{figure*}
	
	Building upon previous research~\cite{vonRohr8465,Guguchia1082}, where a strong pressure effect on superfluid density and a correlation between superfluid density and $T_{\rm c}$ were demonstrated in 2H-NbSe$_2$, there is a pressing need to extend such investigations to encompass other members of the TMD family. Therefore, by systematically probing the superfluid density across different TMDs with and without CDW order we may unravel underlying trends and uncover the interplay between CDW order and superconductivity.
	
	In this letter, we explore the hydrostatic pressure effects on the superfluid density ($n_{s}/m^{*}$) in both 4H-NbSe$_{2}$ and 2H-NbS$_{2}$, where the former exhibits CDW order, whereas the latter does not, while also comparing our findings to 2H-NbSe$_{2}$ with CDW order~\cite{vonRohr8465}. Figure~\ref{Fig1} illustrates the crystal structure of the two studied systems. The 4H and 2H phases of these TMDs differ primarily in their stacking sequences and symmetry. Both phases have the same monolayer. While the 2H phase exhibits a hexagonal structure with AB stacking,  the 4H phase features a more complex hexagonal structure with alternating stacking (ABAC). 
    We observed a 75$\%$ increase in $n_{s}/m^{*}$ in 4H-NbSe$_{2}$ at 2 GPa, while the onset temperature of the CDW order is reduced by only 20$\%$ (from 55 K to 45 K). Notably, this increase in superfluid density in 4H-NbSe$_{2}$ is twice as large as that observed in 2H-NbSe$_{2}$, despite both systems experiencing a similar suppression of CDW order. Additionally, 2H-NbS$_{2}$ also exhibits a 20$\%$ increase in $n_{s}/m^{*}$. Regarding $T_{\rm c}$, it increases by 1 K in 4H-NbSe$_{2}$,  0.5 K in 2H-NbSe$_{2}$, and 0.2 K in 2H-NbS$_{2}$. Therefore, 4H-NbSe$_{2}$ possesses the highest pressure effect on both $n_{s}/m^{*}$ and $T_{\rm c}$ among the three systems studied. Furthermore, we demonstrate an unconventional correlation between $n_{s}/m^{*}$ and $T_{\rm c}$. These findings are discussed in the context of pressure-induced modifications of electron-phonon coupling, $p$-$d$ hybridization, and density of states related to a saddle point situated very close to the Fermi level.

	\begin{figure*}[t!]
		\centering
		\includegraphics[width=\linewidth]{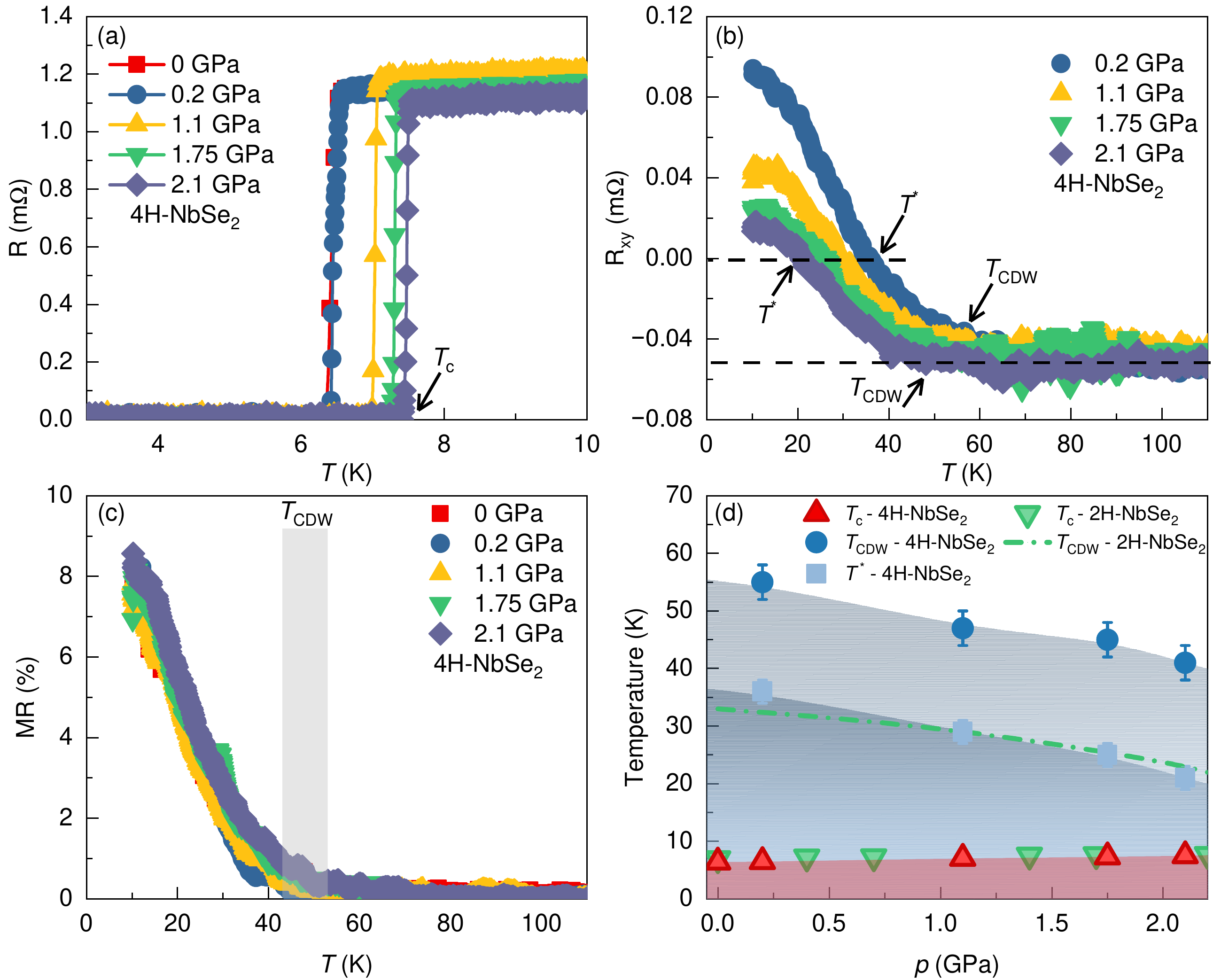}
		\vspace{-0.7cm}
		\caption{\textbf{Temperature-Pressure phase diagram for 4H-NbSe$_{2}$.}
		(a-c) The temperature dependence of (a) longitudinal resistance, (b) the Hall resistance and (c) the magnetoresistance at 9 T for 4H-NbSe$_{2}$, recorded under various hydrostatic pressures. (d) The pressure evolution of the superconducting transition temperature $T_{\rm c}$, the onset of charge density wave transition temperature $T_{\rm CDW}$, and the temperature $T^{*}$ across which the Hall effect changes the sign for 4H-NbSe$_{2}$, obtained from the resistivity measurements. For comparison, the values of $T_{\rm c}$ and $T_{\rm CDW}$ vs pressure for 2H-NbSe$_{2}$ are also shown \cite{vonRohr8465}.}
		\label{Fig1v1}
	\end{figure*}

    First, we present the impact of pressure on superconductivity and charge density wave order in 4H-NbSe$_2$ using magnetotransport measurements. Magnetotransport techniques \cite{Gallo479,Novak041203,Wei015010,Das054702,Li275,LeBoeuf533,LeBoeuf054506}, known for their sensitivity to charge-order transitions, leverage magnetoresistance (MR) as a probe of the mean free path integrated over the Fermi surface \cite{Das054702}. These techniques are particularly effective in detecting changes in scattering anisotropy and Fermi surface reconstructions. The CDW transition is evident in the Hall effect, as previously observed in 2H-NbSe$_2$. Figure 2a illustrates the temperature dependence of the longitudinal resistance measured under varying hydrostatic pressures up to 2.1 GPa. Sharp superconducting transitions are observed, with $T_{\rm c}$ showing a modest increase from 6.3 K at 0 GPa to 7.3 K at 2.1 GPa. Regarding the normal state, we find that the Hall effect exhibits a pronounced temperature dependence below $T_{\rm CDW}$=55 K in 4H-NbSe$_2$, at ambient pressure, consistent with the onset of a CDW. Notably, the Hall signal transitions smoothly from negative (electron-like) to positive (hole-like) across $T^{*}$. This sign reversal parallels observations in 2H-NbSe$_2$	\cite{Li275} and cuprate high-temperature superconductors \cite{LeBoeuf533,LeBoeuf054506}, where it has been attributed to Fermi-surface reconstruction driven by a density-wave phase. Similarly, in 4H-NbSe$_2$, the emergence of a secondary CDW order appears to induce a Fermi-surface reconstruction, resulting in the formation of the hole pocket. Additionally, magnetoresistance in 4H-NbSe$_2$ increases below $T_{\rm CDW}$=55K at ambient pressure, further corroborating the onset of the CDW transition. Under applied pressure, both $T_{\rm CDW}$ and $T^{*}$ exhibit only modest decreases, as summarized in Figure 2(d). For comparison, the pressure dependence of $T_{\rm CDW}$ in 2H-NbSe$_2$ is also shown, highlighting a similar suppression of the CDW transition in both compounds. Within 2.1 GPa, the CDW transition temperature decreases by approximately 20${\%}$, while the superconducting transition temperature increases by 15${\%}$.
		
\begin{figure}[t!]
		\centering
		\includegraphics[width=1.0\linewidth]{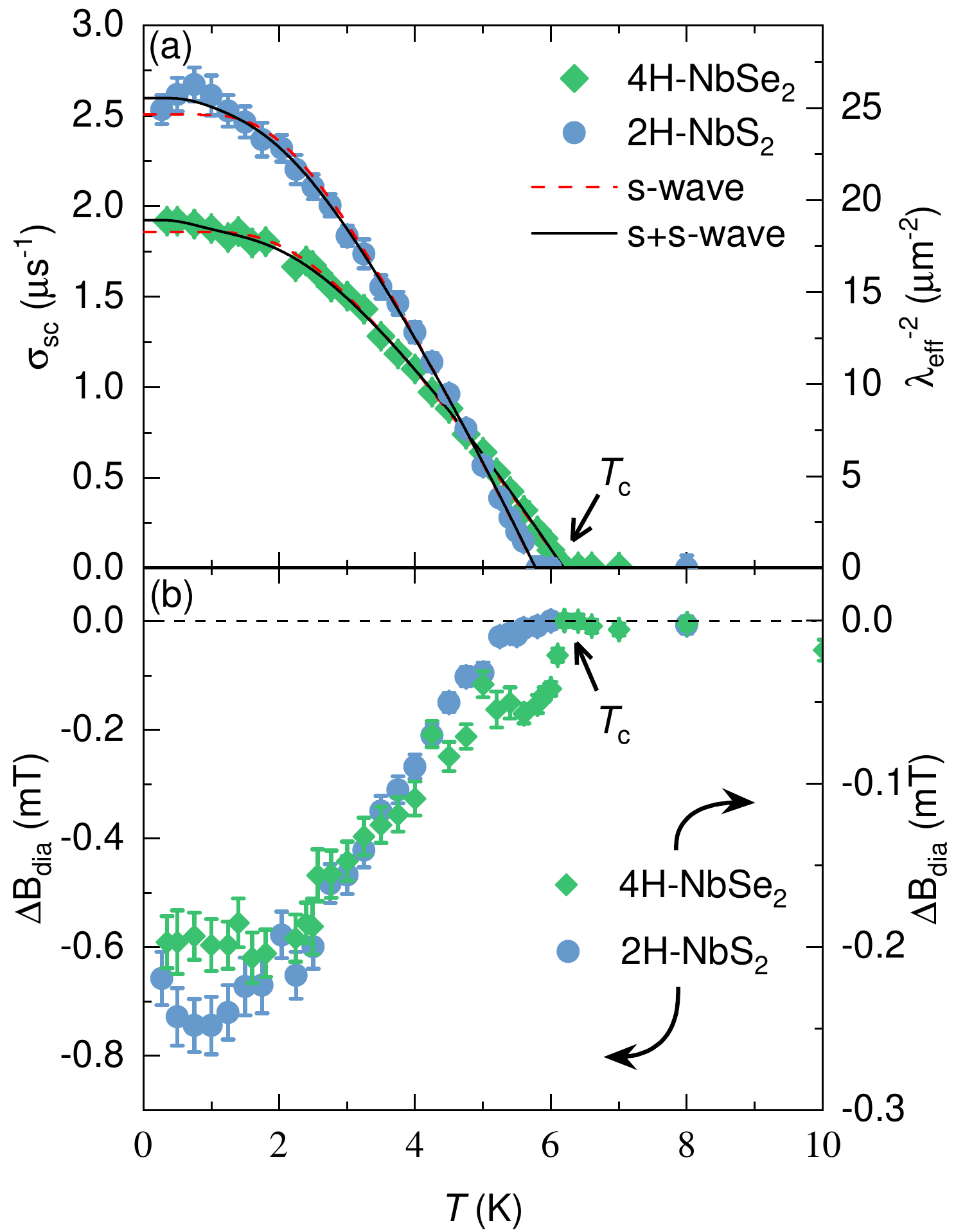}
		\vspace{-0.7cm}
		\caption{\textbf{Temperature evolution of microscopic superconducting quantities.} (a) Temperature dependence of $\sigma_{\rm sc}$ (left $y$-axis) and $\lambda_{\rm eff}^{-2}$ (right $y$-axis) measured at an applied field of $\mu_{\rm 0}H\simeq30$\,mT for both 4H-NbSe$_{2}$  and 2H-NbS$_{2}$. The black solid lines are the fits corresponding to the ($s+s$)-wave model, whereas the red dashed lines correspond to $s$-model fits. (b) Temperature dependence of $\Delta B_{\rm dia}$ calculated from the difference between the internal fields measured at the superconducting and normal state for both systems.}
		\label{Fig2}
	\end{figure}

	Next, we focus on superfluid density measurements for both 4H-NbSe$_2$ and 2H-NbS$_2$ under pressure by examining the superconducting (SC) relaxation rate ($\sigma_{\rm sc}$) in the vortex state. The temperature ($T$) dependence of ($\sigma_{\rm sc}$) is obtained by analyzing (see methods section for details) the TF-$\mu$SR data measured under an applied magnetic field of $\mu_{\rm 0}H\simeq30$\,mT at ambient pressure ($p$) for both 4H-NbSe$_{2}$ and 2H-NbS$_{2}$ and are shown in the left $y$-axis of Fig.~\ref{Fig2}(a). Below $T_{\rm c}$, $\sigma_{\rm sc}$($T$) for both systems gradually increases from zero due to the formation of flux line lattice (FLL) in the SC state and levels off at low temperatures ($T\leq2$\,K). Figure~\ref{Fig2}(b) shows the $T$-dependence of the diamagnetic shifts, $\Delta B_{\rm dia} = B_{{\rm int},s,{\rm sc}} - B_{{\rm int},s,{\rm ns}}$ for both systems. Here, $B_{{\rm int},s,{\rm sc}}$ and $B_{{\rm int},s,{\rm ns}}$ are the local internal fields experienced by the muons at superconducting and normal state of the sample, respectively. $B_{{\rm int},s,{\rm sc}}$ is the calculated from the first moment $\left\langle {B}\right\rangle$ of the asymmetric field distribution $P(B)$ in the superconducting state (see the methods section). As it can be seen from  Fig.~\ref{Fig2}(b), such a strong diamagnetic shift below $T_{\rm c}$ reflects the bulk nature of the superconducting state and rules out any possibility of field induced magnetism in both systems.
	
	The effective (powder average) London magnetic penetration depths ($\lambda_{\rm eff}$) are extracted for both systems from their corresponding $\sigma_{\rm sc}$ by using the relation $\sigma_{\rm sc}(T)/\gamma_{\mu} = 0.06091\Phi_0 \times \lambda_{\rm eff}^{-2}(T)$~\cite{Brandt054506}. Here, ${\gamma_{\mu}}$ is the gyromagnetic ratio of the muon, and $\Phi_0 = 2.068 \times 10^{-15}$\,Wb is the magnetic-flux quantum. $\lambda_{\rm eff}$ is crucial to understanding superconductivity as it is concomitant of superfluid density ($n_{\rm s}/m^{*}$):
	\begin{equation}
		\begin{aligned}
			\sigma_{\rm sc} {\propto} \frac{1}{\lambda_{\rm eff}^2} = \frac{4\pi n_se^2}{m^*c^2} \times  \frac{1}{1 + \xi/l}, 
			\label{eq1}
		\end{aligned}
	\end{equation}
	where $m^{*}$ is the effective mass of superconducting carriers, ${\xi}$ and $l$ are the coherence length and mean free path, respectively. In case of systems close to the clean limit (${\xi}$/$l$ ${\rightarrow}$ 0), $\lambda_{\rm eff}^{-2}$ is directly proportional to the superfluid density, $\lambda_{\rm eff}^{-2}\propto n_{s}/m^{*}$ as the second term in Eq.~\eqref{eq1} essentially becomes unity. By using the Ginzburg-Landau theory ($\xi$ =  $\sqrt{\Phi_0/2\pi\,H_{c2}}$) and reported upper critical field ($H_{c2}$) values for 4H-NbSe$_{2}$ [$\mu_{0}H_{c2}(0) = 26.5$\,T]~\cite{Zhou224518} and for 2H-NbS$_{2}$ [$\mu_{0}H_{c2}(0)\|_{\rm ab} = 23$\,T]~\cite{Cho3676}, we have estimated the coherence lengths to be $\xi\simeq3.5$\,nm and $\xi\|_{\rm ab}\simeq3.8$\,nm for 4H-NbSe$_{2}$ and 2H-NbS$_{2}$, respectively, at ambient pressure. Although no accurate estimate for $l$ is available, it is reasonable to assume a very little effect from ${\xi}$/$l$ in view of their very small $\xi$. Thus we can reliably take $\sigma_{\rm sc}$ and/or $\lambda_{\rm eff}^{-2}$ as a measure of superfluid density ($\sigma_{\rm sc} \propto \lambda_{\rm eff}^{-2} \propto n_{\rm s}/m^*$) as both systems lie close to the clean limit. To give a quantitative idea of $n_{\rm s}/m^*$, we show $\lambda_{\rm eff}^{-2}$ in the right $y$-axis of Fig.~\ref{Fig2}(a).
	
	${\sigma}_{{\rm sc}}$($T$) is sensitive to the topology of the superconducting gap. In our case, the observed  low-$T$ leveling off behavior is consistent with the scenario of nodeless superconductivity, for which  $\lambda_{\rm eff}^{-2}(T)$ approaches the zero-$T$ value exponentially. To unveil the nature of the superconducting gap structure in 4H-NbSe$_{2}$  and 2H-NbS$_{2}$, the $T$-dependent behavior of $\sigma_{\rm sc}$ for both compounds are analyzed by the empirical ${\alpha}$-model~\cite{Guguchia3672} within local (London) approximation (${\lambda}$ ${\gg}$ ${\xi}$). The main assumption of this model is that the gaps appearing in different bands are independent from one another, despite having a common $T_{\rm c}$. So, the superfluid densities for each component can be calculated (for details see the methods section) independently and added together with their respective weight factors as shown below:
	\begin{equation}
		\frac{\sigma_{\rm sc}(T)}{\sigma_{\rm sc}(0)}=\omega\frac{\sigma_{\rm sc}(T,\Delta_{0,1})}{\sigma_{\rm sc}(0,\Delta_{0,1})}+ (1-\omega)\frac{\sigma_{\rm sc}(T,\Delta_{0,2})}{\sigma_{\rm sc}(0,\Delta_{0,2})}.
	\end{equation}
	
	$\sigma_{\rm sc}(0) \propto \lambda_{\rm eff}^{-2} (0)$ and $\lambda_{\rm eff}(0)$ is the magnetic penetration depth at $T=0$~K. ${\Delta_{0,i}}$ measures the value of $i$-th ($i=1$, 2) SC gap at $T=0$~K, whilst $\omega$ and ($1-\omega$) are the weight factors quantifying their relative contributions to $\sigma_{\rm sc}$. The results of our analysis are shown in Fig.~\ref{Fig2}(a), in which a single $s$-wave (red dashed line) and a ($s+s$)-wave (black solid line) models are considered. It is clearly evident that a two gap ($s+s$)-wave model describes our data much better than a single gap $s$-wave model. 
Two-gap superconductivity in both systems can be understood by assuming that the SC gaps open at two distinct types of bands crossing the Fermi level ($E_{\rm F}$). The values of superconducting gaps ${\Delta}_{1}$ and ${\Delta}_{2}$ (see Fig. 8 for the pressure evolution of the SC gaps) are in good agreement with previous reports~\cite{vonRohr8465,Zhou224518,Diener054531}.	
	
	
	\begin{figure*}[t!]
		\centering
		\includegraphics[width=1.0\linewidth]{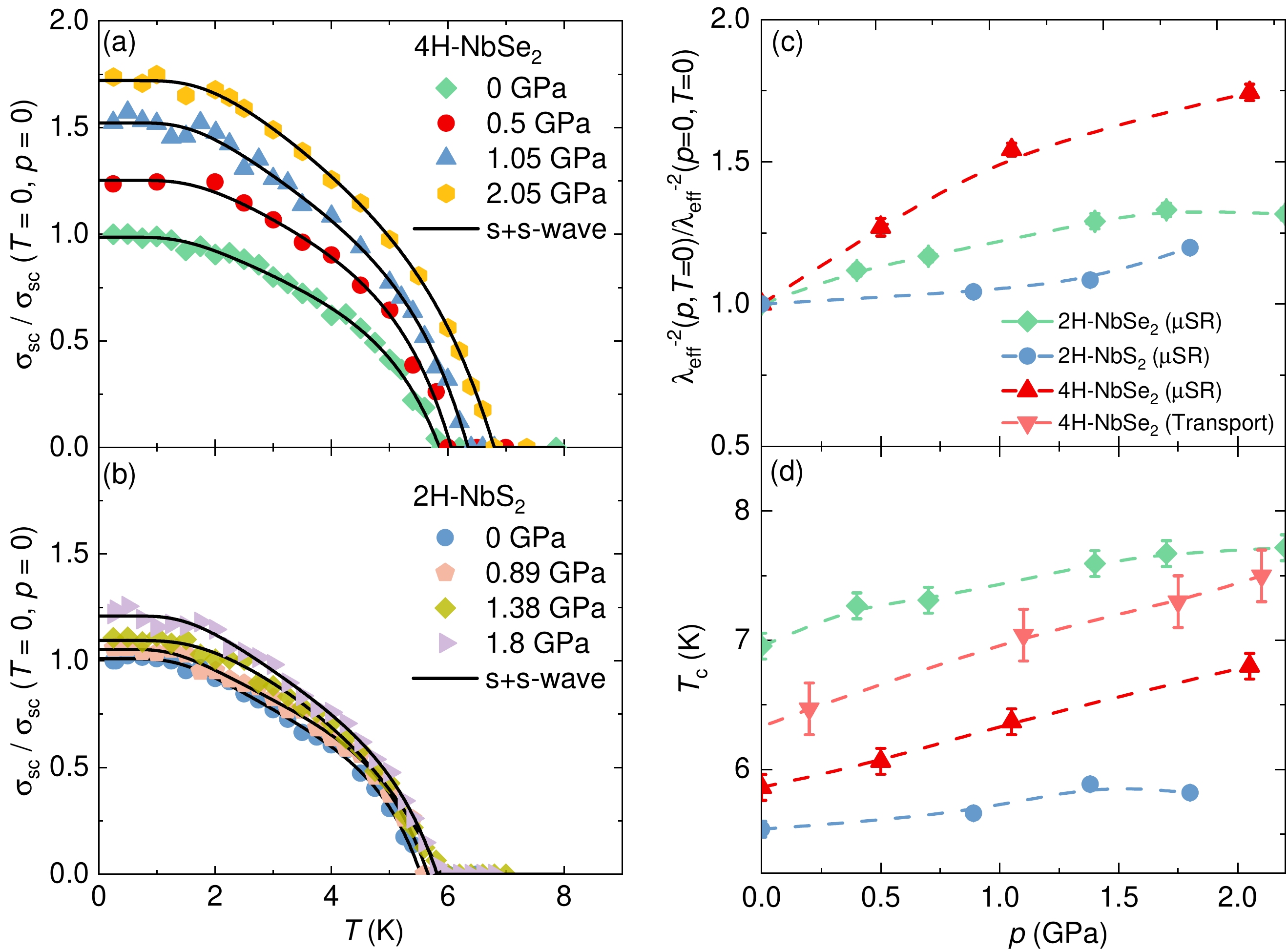}
		\vspace{-0.7cm}
		\caption{\textbf{Temperature and pressure evolution of superconducting quantities.}
		The temperature dependence of $\sigma_{\rm sc} (T)$ (normalized to the value at zero applied pressure) under different applied hydrostatic pressures for (a) 4H-NbSe$_{2}$ and (b) 2H-NbS$_{2}$, respectively. The solid lines are the fits corresponding to ($s+s$)-wave model. Pressure dependence of (c) normalized zero temperature value of $\lambda_{\rm eff}^{-2} (0)$ and (d) $T_{\rm c}$. The values of $T_{\rm c}$ for 4H-NbSe$_{2}$, determined from both resistivity and $\mu$SR experiments, are shown.} 
		\label{Fig3}
	\end{figure*}
	
	In order to study the pressure evolution of the superconducting order parameters of 4H-NbSe$_{2}$ and 2H-NbS$_{2}$, TF-$\mu$SR experiments are carried out under pressure range of $0\leq p \leq2$~GPa. The TF-$\mu$SR time spectra recorded at different hydrostatic pressures are analyzed by adopting the procedure described in the methods section (see Fig. 7a-d). To highlight the relative change of the superfluid density with pressure, we present the extracted $T$-dependence of $\sigma_{\rm sc}$ at various applied pressures after normalizing each one to the value obtained at zero pressure and temperature for 4H-NbSe$_{2}$ and 2H-NbS$_{2}$ in Fig.~\ref{Fig3}(a) and (b), respectively.
	The two gap ($s+s$)-wave model adequately describes all the experimental data up to the highest applied pressures with nearly pressure independent weight factors of $\omega \sim$\,0.54 and 0.52 for 4H-NbSe$_{2}$ and 2H-NbS$_{2}$, respectively. The corresponding fits are shown by the black solid lines in Fig.~\ref{Fig3}(a) and (b). The absolute values of the two superconducting gaps (${\Delta}_{1}$ and ${\Delta}_{2}$) extracted from the ($s+s$)-wave model fits remain almost constant throughout the applied pressure range for both systems, as depicted in Fig.~8 of methods section. The pressure evolution of the other extracted superconducting parameters such as $\lambda_\mathrm{eff}^{-2} (0) [\propto n_{s}/m^{*}]$ and $T_{\rm c}$ are plotted in Fig.~\ref{Fig3}(c) and (d), respectively, along with the corresponding pressure dependent behavior of 2H-NbSe$_{2}$ for comparison, taken from our recent report, see Ref.~\cite{vonRohr8465}. As it is evident from Fig.~\ref{Fig3}(c), the relative increase of $n_{s}/m^{*}$ with pressure for 4H-NbSe$_{2}$ is found to be the highest, and corresponds to a 75$\%$ increase at the maximum applied pressure of 2.05~GPa. The relative increase is of 32 $\%$ for 2H-NbSe$_{2}$ and only 20 $\%$ for 2H-NbS$_{2}$, at the maximum applied pressures of 2.2~GPa and 1.8~GPa, respectively. However, the critical temperatures $T_{\rm c}$ of all systems are found to be notably less influenced by the increase of pressure. $T_{c}$ increases by 0.94~K (16\%) for 4H-NbSe$_{2}$, 0.76~K (11\%) for 2H-NbSe$_{2}$ and 0.28~K (5\%) for 2H-NbS$_{2}$.
	
The primary discovery of this study lies in the remarkable increase of superfluid density ($n_{s}/m^{*}$) under pressure, particularly in 4H-NbSe$_{2}$, surpassing its counterparts 2H-NbSe$_{2}$ and 2H-NbS$_{2}$. The $p$-$T$ phase diagram of 2H-NbSe$_{2}$, which has been thoroughly explored in the past ~\cite{Berthier1393, Suderow117006, Feng7224}, indicates a linear rise in $T_{\rm c}$ with pressure, alongside to a systematic suppression of the CDW order ($T_{\rm CDW}$). Here, we have constructed the temperature-pressure phase diagram for 4H-NbSe$_{2}$  for the first time using magnetotransport measurements, revealing a similar increase in $T_{\rm c}$ and a comparable suppression of the CDW onset temperature as observed in 2H-NbSe$_{2}$. The antagonistic pressure dependence of $T_{\rm c}$ and $T_{\rm CDW}$ sparked the notion of mutual competition between CDW order and superconductivity, which continues to be an ongoing debate~\cite{Moulding043392,Cao245125}. However, despite the absence of a CDW state in all known NbS$_{2}$ polytypes, the $T_{\rm c}$ of 2H-NbS$_{2}$ is comparable to that of 4H-NbSe$_{2}$, albeit with a higher superfluid density. Additionally, the modest suppression of the CDW state within the 2.1 GPa pressure range for 2H-NbSe$_2$, and 4H-NbSe$_2$, suggests that the substantial increase in superfluid density cannot solely stem from the CDW state suppression, especially considering the maximal availability of the density of states, $\Delta D(E_{\rm F}) \approx$ 1 $\%$~\cite{Rossnagel235119,Johannes205102}. Remarkably, a 20$\%$ increase in the superfluid density is also observed for 2H-NbS$_{2}$, which is devoid of the CDW order present in the other TMDs (4H/2H)-NbSe$_{2}$. The insensitivity of the two superconducting gaps ($\Delta_{1}$, $\Delta_{2}$) to pressure changes across all investigated TMDs (4H/2H)-NbSe$_{2}$ with and 2H-NbS$_{2}$~\cite{Tissen134502,Heil087003,Youbi155105} without a CDW suggests a more complex origin for the significant pressure effects beyond competing CDW order.

	\begin{figure}[t!]
		\centering
		\includegraphics[width=1.0\linewidth]{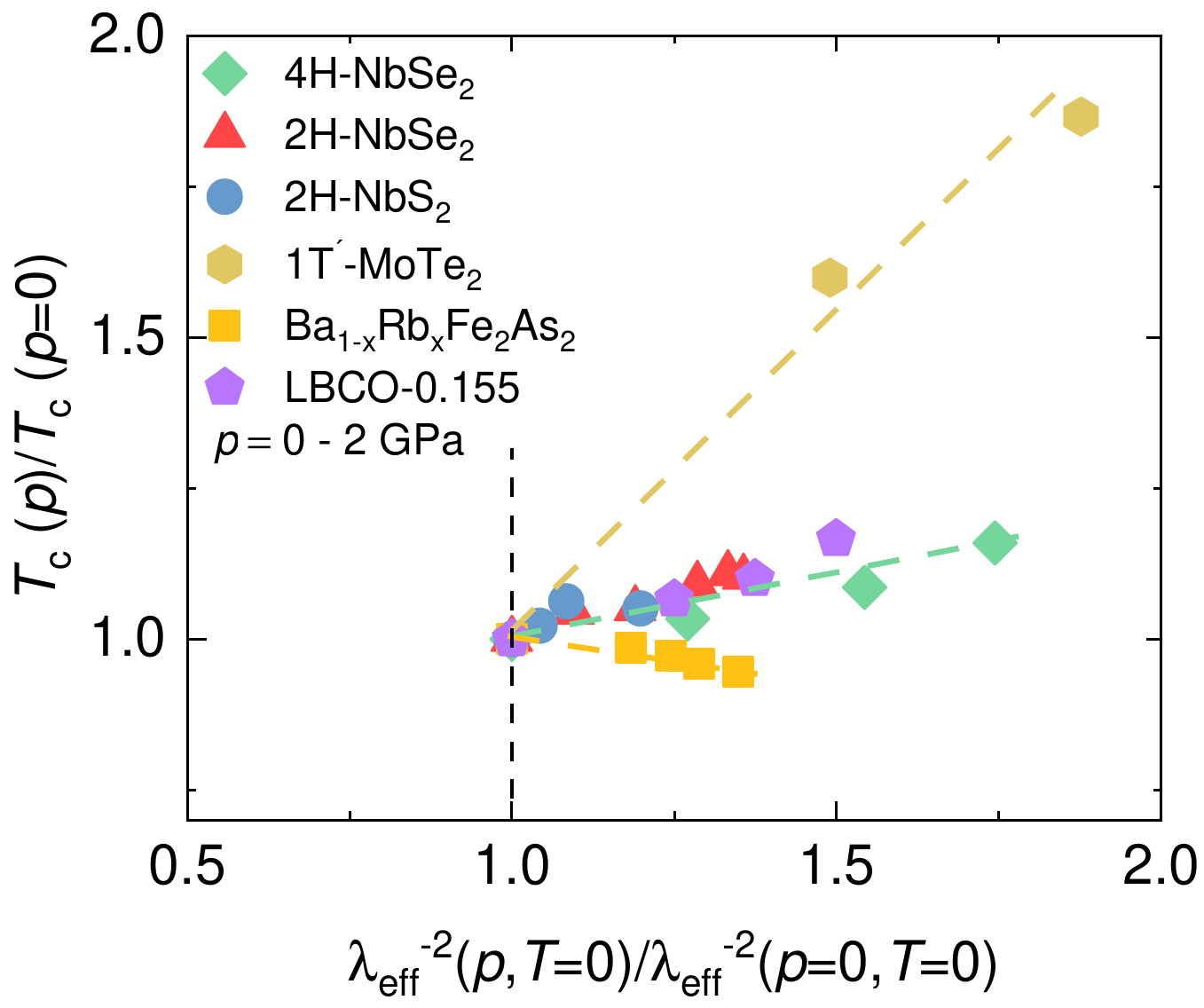}
		\vspace{-0.7cm}
		\caption{\textbf{Superconducting critical temperature versus Superfluid Density.} 
		Plot of $T_{\rm c}(p)$/$T_{\rm c}(p=0)$ vs normalized superfluid density at different applied pressures obtained from our ${\mu}$SR experiments in 4H-NbSe$_{2}$, 2H-NbSe$_{2}$ \cite{vonRohr8465}, 2H-NbS$_{2}$, 1T$'$-MoTe$_{2}$ \cite{Guguchia1082}, Ba$_{1-x}$Rb$_{x}$Fe$_{2}$As$_{2}$ \cite{Guguchia8863}, and La$_{2-x}$Ba$_{x}$CuO$_{4}$ ($x$ = 0.155) \cite{Guguchia214511}. Black dashed vertical line manifests the BCS expectation.}
		\label{Fig4}
	\end{figure}
    
In 2H-NbSe$_{2}$, the Fermi surface (FS) structure is intricate, featuring predominantly four Nb ($4d$)-orbital derived bands forming cylindrical Fermi surface sheets centered around the mid-$\Gamma$ and corner-${\rm K}$ points, along with one Se ($4p$)-orbital derived band resembling a small 3D "pancake" FS centered around the $\Gamma$-point. Superconductivity manifests through the opening of two gaps: a relatively large gap spanning most of the Nb-derived FS sheets and a smaller one at the Se-derived band. Meanwhile, a coexisting CDW opens a gap on a small portion of the Nb-derived FS sheets, distinct from the superconducting gaps in $k$-space~\cite{Yokoya2518, Borisenko166402, Rahn224532}. There is a growing evidence that the CDW order in 2H-NbSe$_{2}$ is primarily driven by momentum ($q$)-dependent electron-phonon coupling (EPC) rather than the previous belief of mechanisms involving Fermi-surface nesting and saddle-points etc.~\cite{Johannes205102, Johannes165135, Weber107403, Valla086401}. Furthermore, recent $ab$-initio band structure calculations on 4H-NbSe$_{2}$ \cite{Zhou224518} propose that the lower $T_{\rm c}$ and higher $T_{\rm CDW}$ could be linked to weaker $p$-$d$ hybridization and the availability of additional density of states near the Fermi level compared to 2H-NbSe$_{2}$. Consequently, a pressure-dependent complex interplay among various factors such as EPC, multiband FS structure, and $p$-$d$ hybridization likely underlies the substantial pressure-induced change in superfluid density observed in 4H-NbSe$_{2}$~\cite{Suderow117006, Leroux140303}.

The strong increase in superfluid density in TMDs underscores their unconventional superconductivity, contrasting with the behavior observed in conventional BCS superconductors. This is supported by the correlation between $T_{\rm c}$ and zero-$T$ value of the superfluid densities ($\lambda_{\rm eff}^{-2}$(0)) observed in 4H-NbSe$_{2}$, 2H-NbS$_{2}$, as well as previously reported materials such as 2H-NbSe$_{2}$~\cite{vonRohr8465} and 1T$'$-MoTe$_{2}$~\cite{Guguchia1082}. The plotted data in Fig.~\ref{Fig4}, illustrating the relative change in $T_{\rm c}$ as a function of the relative change in superfluid density~\cite{Das217002}, shows that the slopes for 2H-NbSe$_{2}$,  4H-NbSe$_{2}$ and 2H-NbS$_{2}$ are the same, suggesting a similar mechanism for the pressure-induced enhancement of the superfluid density. Intriguingly, this slope resembles that observed in optimally doped cuprates~\cite{Guguchia214511} and Fe-based superconductors ~\cite{Guguchia8863}, although the sign of the slope is opposite in Fe-based systems. Conversely, 1T$'$-MoTe$_{2}$ exhibits a stronger slope, despite its $T_{\rm c}$ remaining below the optimal value even under pressure~\cite{Guguchia1082,Qi11038}. In contrast, 4H-NbSe${2}$, 2H-NbSe$_{2}$~\cite{vonRohr8465}, and 2H-NbS$_{2}$ are closer to their optimal $T_{\rm c}\simeq$ 8 K~\cite{Majumdar084005}.

The observed correlation between $T_{\rm c}$ and the superfluid density, initially observed in cuprates~\cite{Uemura2317} and later extended to Fe-based unconventional superconductors~\cite{Luetkens305,Khasanov092506,Luetkens097009} and kagome-lattice superconductors \cite{Guguchia41}, reveals intriguing insights~\cite{Das217002}. In cuprates, $T_{\rm c}$ is about 4-5 times lower than the expected ideal Bose Condensation temperature ($T_{\rm B}$) for a non-interacting Bose gas, indicating an evolution from unconventional Bose-Einstein condensation (BEC)-like to conventional BCS-type condensation scenario ~\cite{Uemura2665,Uemura605,UemuraS4515}. In the investigated TMDs, the ratio $T_{\rm c}$/$T_{\rm F}$ is reduced by approximately 20 times compared to cuprates, yet the unconventional correlation persists. This striking similarity across distinct classes of superconductors implies a shared mechanism underlying superconductivity and other quantum orders in these materials.

In conclusion, we investigated the effects of hydrostatic pressure on the microscopic properties of superconductivity in 4H-NbSe$_{2}$ and 2H-NbS$_{2}$, alongside the pressure evolution of the CDW in 4H-NbSe$_{2}$ using magnetotransport and ${\mu}$SR experiments. We find that the CDW onset temperature in 4H-NbSe$_{2}$ decreases by only 20${\%}$ (from 55 K to 45 K), similar to 2H-NbSe$_{2}$. Despite similar CDW suppression in 2H- and 4H-NbSe$_{2}$, the superfluid density increase is twice as strong in 4H-NbSe$_{2}$. Additionally, a sizeable increase in superfluid density is observed in 2H-NbS$_{2}$, which lacks CDW order. These findings lead us to infer that the increase in the superfluid density and the underlying superconductivity in these systems are minimally impacted by pressure-induced changes in CDW order. Instead, they may have a more complex origin, possibly involving additional factors beyond the influence of CDW order. Notably, an unconventional correlation  between $T_{c}$ and $n_{s}/m^{*}$ is evident across all the studied TMDs, including the previously investigated topological superconductor 1T$'$-MoTe$_{2}$, reflecting a characteristic feature of unconventional superconductors, such as cuprates and Fe-based superconductors. We anticipate that our study will inspire future theoretical investigations, such as band structure calculations under pressure, as well as experimental investigations (e.g., ARPES), to achieve a comprehensive understanding of superconductivity in these TMDs, particularly in 4H-NbSe$_{2}$.
	
	\section{Acknowledgments}~
	The ${\mu}$SR experiments were carried out at the Swiss Muon Source (S${\mu}$S) Paul Scherrer Insitute, Villigen, Switzerland. This work was supported by the Swiss National Science Foundation (SNSF) through the SNSF Starting Grant No. TMSGI2${\_}$211750.\\ 

    \section{Author contributions}~
Z.G. conceived and designed the project. Z.G. and F.v.R. supervised the project. Crystal growth: C.W. and F.v.R.. High pressure magnetotransport experiments and analysis: V.S. and Z.G. with contributions from O.G., P.K. and M.B.. Muon spin rotation experiments under pressure and corresponding analysis: Z.G., S.S.I., J.N.G., H.L., and R.K.. Figure development and writing of the paper: Z.G. and S.S.I. with contributions from all authors. All authors discussed the results, interpretation, and conclusion.\\

\section{METHODS}

\textbf{Sample preparation}: Single-phase polycrystalline samples of 2H-NbS$_2$ and 4H-NbSe$_2$ were prepared by means of high-temperature solid-state synthesis, described elsewhere~\cite{vonRohr8465}.\\

\textbf{Pressure cells for $\mu$SR and magnetotransport experiments}: A double-wall piston-cylinder cell, constructed from CuBe material and specifically designed for ${\mu}$SR experiments under pressure, was utilized to generate pressures up to 2.2 GPa \cite{Khasanov140,Khasanov190903}. Daphne oil served as the pressure-transmitting medium. The pressure calibration was established by measuring the superconducting transition of a small indium plate inside the sample volume from AC susceptibility. The pressure cell was optimized for maximum filling factor. Approximately 40${\%}$ of the muons stopped within the sample during the experiments.\\ 

Magnetotransport measurements were performed using a double-wall piston-cylinder cell with a BeCu (Beryllium Copper) outer body and a Super-Alloy inner body, integrated into a Quantum Design PPMS. The pressure at 300 K was determined using the resistance of Manganin, while the low-temperature pressure was calibrated by monitoring the superconducting transition of Sn. Daphne oil was employed as the pressure-transmitting medium.\\

\textbf{$\mu$SR experiment}: In a ${\mu}$SR experiment nearly 100 ${\%}$ spin-polarized muons $\mu$$^{+}$ are implanted into the sample one at a time. The positively charged $\mu$$^{+}$ thermalize at interstitial lattice sites, where they act as magnetic microprobes. In a magnetic material the muon spin precesses in the local field $B_{\rm \mu}$ at the muon site with the Larmor frequency ${\nu}_{\rm \mu}$ = $\gamma_{\rm \mu}$/(2${\pi})$$B_{\rm \mu}$ (muon gyromagnetic ratio $\gamma_{\rm \mu}$/(2${\pi}$) = 135.5 MHz T$^{-1}$). Using the $\mu$SR technique important length scales of superconductors can be measured, namely the magnetic penetration depth $\lambda$ and the coherence length $\xi$. If a type II superconductor is cooled below $T_{\rm c}$ in an applied magnetic field ranged between the lower ($H_{c1}$) and the upper ($H_{c2}$) critical fields, a vortex lattice is formed which in general is incommensurate with the crystal lattice with vortex cores separated by much larger distances than those of the unit cell. Because the implanted muons stop at given crystallographic sites, they will randomly probe the field distribution of the vortex lattice. Such measurements need to be performed in a field applied perpendicular to the initial muon spin polarization (so called TF configuration). 

$\mu$SR experiments under pressure were performed at the ${\mu}$E1 beamline of the Paul Scherrer Institute (Villigen, Switzerland, where an intense high-energy ($p_{\mu}$ = 100 MeV/c) beam of muons is implanted in the sample through the pressure cell. The low background Dolly instrument was used to study the polycrystalline samples of 2H-NbSe$_{2}$ and 4H-NbSe$_{2}$ at ambient pressure.\\

\textbf{Analysis of TF-${\mu}$SR data}: Figure~\ref{FigS2}(a) and (c) depict the transverse-field (TF) ${\mu}$SR time spectra measured at two selected temperatures above (10\,K) and below (0.3\,K) $T_{\rm c}$ under an applied magnetic field of $\mu_{\rm 0}H \simeq 30$\,mT for 4H-NbSe$_{2}$  and 2H-NbS$_{2}$, respectively. At 10\,K, the oscillations show a very small relaxation due to the random local fields produced by the nuclear magnetic moments in the normal state (NS). A significant increase in the relaxation rate is observed at 0.3\,K as muons sense an inhomogeneous distribution of the local magnetic fields generated by the flux-line-lattice (FLL) formed in the superconducting (SC) state below $T_{\rm c}$. The normalized Fourier transforms of the TF- ${\mu}$SR time spectra are depicted in Fig.~\ref{FigS2}(b) and (d) for 4H-NbSe$_{2}$  and 2H-NbS$_{2}$, respectively. The formation of FLL is further evident from the asymmetric broad Fourier spectra observed at 0.3\,K, while at 10\,K typical symmetric sharp Fourier spectra (centered around 30\,mT) are observed for both 4H-NbSe$_{2}$  and 2H-NbS$_{2}$. 

In order to model such broad asymmetric field distribution [$P (B)$] in the SC state, the ambient pressure TF-${\mu}$SR time spectra measured below $T_{\rm c}$ for all samples are analyzed by using the following two-component functional form

\begin{equation}
	\begin{aligned}
		A_{\rm TF} (t) = \sum_{i=0}^{2} A_{s,i}\exp\Big[-\frac{\sigma_{i}^2t^2}{2}\Big]\cos(\gamma_{\mu}B_{{\rm int},s,i}t+\varphi)  
	\end{aligned}
\end{equation}

Here $A_{s,i}$, $\sigma_{i}$ and $B_{{\rm int},s,i}$ is the initial asymmetry, relaxation rate and local internal magnetic field of the $i$-th component. ${\varphi}$ is the initial phase of the muon-spin ensemble. $\gamma_{\mu}/(2{\pi})\simeq 135.5$~MHz/T is the muon gyromagnetic ratio. The first and second moment of the local magnetic field distribution are give by~\cite{Khasanov104504}

\begin{equation}
	\begin{aligned}
		\left\langle {B}\right\rangle = \sum_{i=0}^{2} \frac {A_{s,i}B_{{\rm int},s,i}}{A_{s,1}+A_{s,2}}
		\label{eqS2}
	\end{aligned}
\end{equation}
and 
\begin{equation}
	\begin{aligned}
		\left\langle {\Delta B}\right\rangle ^2 = \frac {\sigma ^2}{\gamma_{\mu}^2} =  \sum_{i=0}^2  \frac {A_{s,i}}{A_{s,1}+A_{s,2}} \Big[\sigma_i^2/\gamma_{\mu}^2 + \left(B_{{\rm int},s,i} - \langle B \rangle\right)^2\Big].
	\end{aligned}
\end{equation}

Above $T_{\rm c}$, in the normal state, the symmetric field distribution could be nicely modeled by only one component. The obtained relaxation rate and internal magnetic field are denoted by $\sigma_{\rm ns}$ and $B_{{\rm int},s,{\rm ns}}$. $\sigma_{\rm ns}$ is found to be small and temperature independent (dominated by nuclear magnetic moments) above $T_{\rm c}$ and assumed to be constant in the whole temperature range. Below $T_{\rm c}$, in the SC state the relaxation rate and internal magnetic field are indicated by $\sigma_{\rm sc}$ and $B_{{\rm int},s,{\rm sc}}$. $\sigma_{\rm sc}$ is extracted by using $\sigma_\mathrm{sc} = \sqrt{\sigma^{2} - \sigma^{2}_\mathrm{ns}}$. $B_{{\rm int},s,{\rm sc}}$ is evaluated from $\left\langle {B}\right\rangle$ using Eq.~\eqref{eqS2}.

The TF ${\mu}$SR time spectra obtained from GPD spectrometer are analyzed by using the following functional form:

\begin{equation}
	\begin{aligned}
		A_{\rm TF} (t) = A_s\exp\Big[-\frac{(\sigma_{{\rm sc}}^2+\sigma_{{\rm ns}}^2)t^2}{2}\Big]\cos(\gamma_{\mu}B_{int,s}t+\varphi) \\
		+ A_{pc}\exp\Big[-\frac{\sigma_{pc}^2t^2}{2}\Big]\cos(\gamma_{\mu}B_{int,pc}t+\varphi), 
		\label{eqS4}
	\end{aligned}
\end{equation}

Here $A_{\rm s}$ and $A_{\rm pc}$  denote the initial assymmetries of the sample and the pressure cell, respectively. In the analysis ${\sigma}_{\rm ns}$ was assumed to be constant over the entire temperature range and was fixed to the value obtained above $T_{\rm c}$ where only nuclear magnetic moments contribute to the muon depolarization rate ${\sigma}$. The Gaussian relaxation rate, ${\sigma}_{\rm pc}$, reflects the depolarization due to the nuclear moments of the pressure cell. The width of the pressure cell signal increases below $T_{c}$. As shown previously \cite{Maisuradze184523}, this is due to the influence of the diamagnetic moment of the SC sample on the pressure cell, leading to the temperature dependent ${\sigma}_{\rm pc}$ below $T_{c}$. In order to consider this influence we assume the linear coupling between ${\sigma}_{\rm pc}$ and the diamagnetic field shift the SC state:\\
${\sigma}_{\rm pc}$($T$) = ${\sigma}_{\rm pc}$($T$ ${\textgreater}$ $T_{\rm c}$) + $C(T)$($\Delta B_{\rm dia}$), where  ${\sigma}_{\rm pc}$($T$ ${\textgreater}$ $T_{\rm c}$) = 0.25 ${\mu}$$s^{-1}$ is the temperature independent Gaussian relaxation rate. $\Delta B_{\rm dia} = B_{{\rm int},s,{\rm sc}} - B_{{\rm int},s,{\rm ns}}$,  $B_{{\rm int},s,{\rm ns}}$ and $B_{{\rm int},s,{\rm sc}}$ are the internal magnetic fields measured in the normal and in the SC state, respectively.\\ 

\textbf{Analysis of ${\lambda}(T)$}: ${\lambda}$($T$) was calculated within the local (London) approximation (${\lambda}$ ${\gg}$ ${\xi}$) by the following expression \cite{Suter69, Tinkham2004}:
\begin{equation}
	\frac{\sigma_{\rm sc}(T,\Delta_{0,i})}{\sigma_{\rm sc}(0,\Delta_{0,i})}=
	1+\frac{1}{\pi}\int_{0}^{2\pi}\int_{\Delta(_{T,\varphi})}^{\infty}(\frac{\partial f}{\partial E})\frac{EdEd\varphi}{\sqrt{E^2-\Delta_i(T,\varphi)^2}},
\end{equation}
where $f=[1+\exp(E/k_{\rm B}T)]^{-1}$ is the Fermi function, ${\varphi}$ is the angle along the Fermi surface, and ${\Delta}_{i}(T,{\varphi})={\Delta}_{0,i}{\Gamma}(T/T_{\rm c})g({\varphi}$)
(${\Delta}_{0,i}$ is the maximum gap value at $T=0$). The temperature dependence of the gap is approximated by the expression ${\Gamma}(T/T_{\rm c})=\tanh{\{}1.82[1.018(T_{\rm c}/T-1)]^{0.51}{\}}$,\cite{Carrington205} while $g({\varphi}$) describes the angular dependence of the gap and it is replaced by 1 for both an $s$-wave and an $s$+$s$-wave gap, and ${\mid}\cos(2{\varphi}){\mid}$ for a $d$-wave gap~\cite{Guguchia8863}.\\
%

\bibliography{reff_NbS2}

\begin{figure*}[t!]
	\centering
	\includegraphics[width=1.0\linewidth]{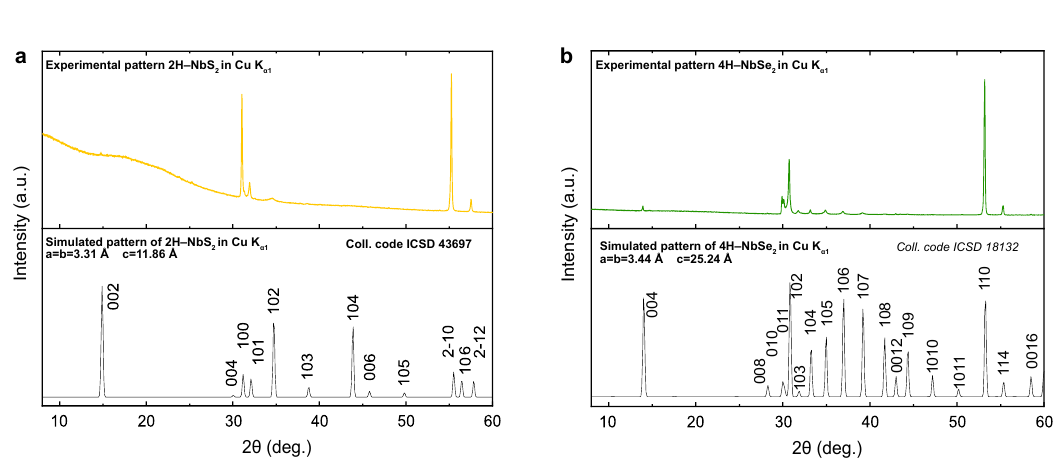}
	\vspace{-0.3cm}
	\caption{(a): Top: Experimental pattern of 2H-NbS$_2$ in yellow. Bottom: Simulated PXRD patterns of 2H-NbS$_2$ from Ref.~\cite{Jellinek376}, including the respective Miller indices of the reflexions. (b) Top: Experimental pattern of 4H-NbSe$_2$ in green. Bottom: Simulated PXRD patterns of 4H-NbSe$_2$ from Ref.~\cite{Brown31}, including the respective Miller indices of the reflexions. The experimental patterns show strong preferred orientation of the (hk0) reflections due to the transmission geometry of the measurement.}
	\label{FigS1}
\end{figure*}

\begin{figure*}[t!]
	\centering
	\includegraphics[width=1.0\linewidth]{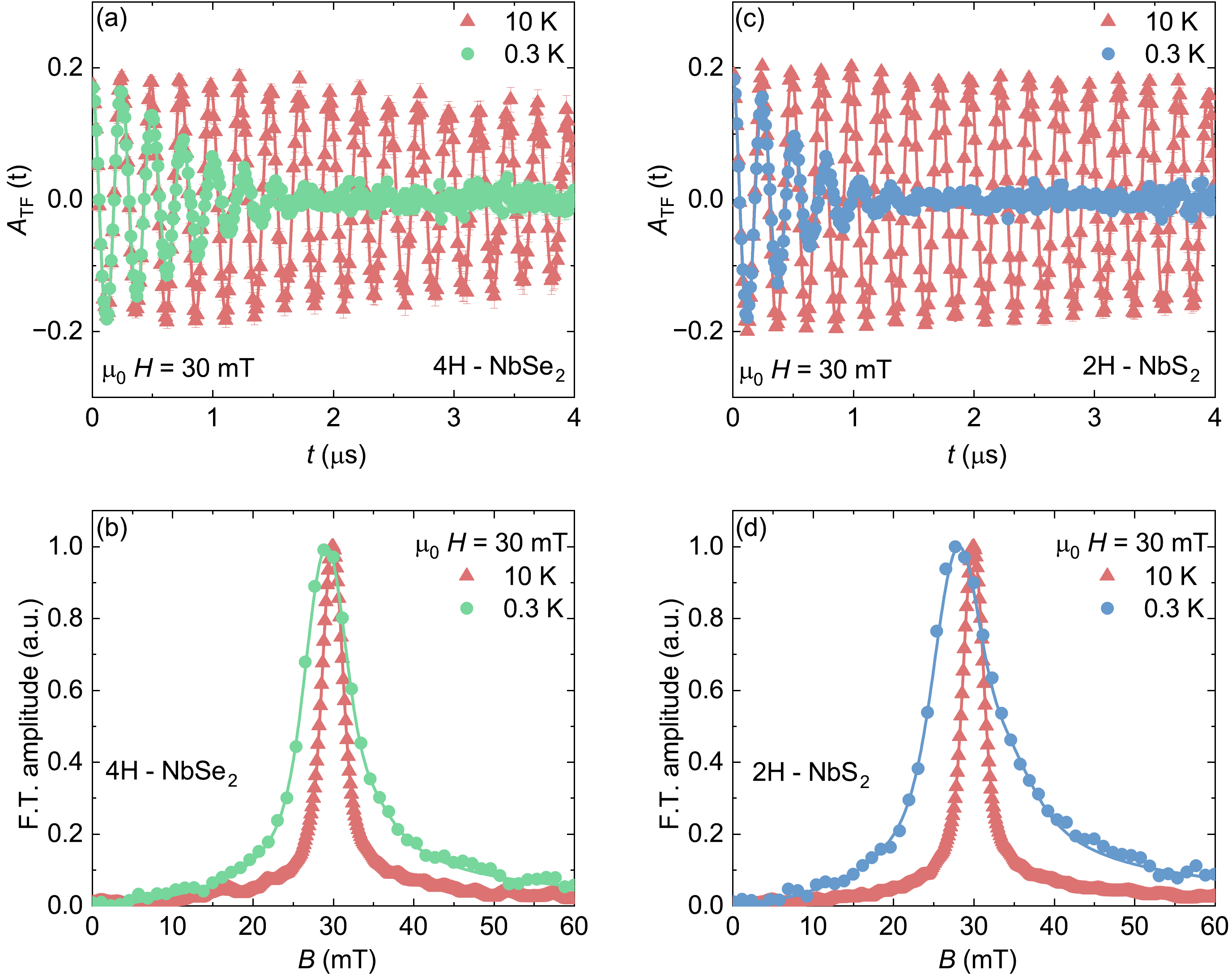}
	\vspace{0cm}
	\caption{\textbf{Transverse-field (TF) $\mu$SR time spectra and the corresponding normalized Fourier spectra} TF-$\mu$SR time spectra recorded above and below $T_{\rm c}$ after field cooling the sample from  above $T_{\rm c}$ under an applied magnetic field of 30~mT for (a) 4H-NbSe$_{2}$  and (c) 2H-NbS$_{2}$, respectively. Solid lines correspond to the fits as described in the text. Corresponding normalized Fourier transforms of the TF-$\mu$SR time spectra and associated fittings are shown in (b) for 4H-NbSe$_{2}$  and in (d) for 2H-NbS$_{2}$.}
	\label{FigS2}
\end{figure*}

\begin{figure*}[t!]
	\centering
	\includegraphics[width=1.0\linewidth]{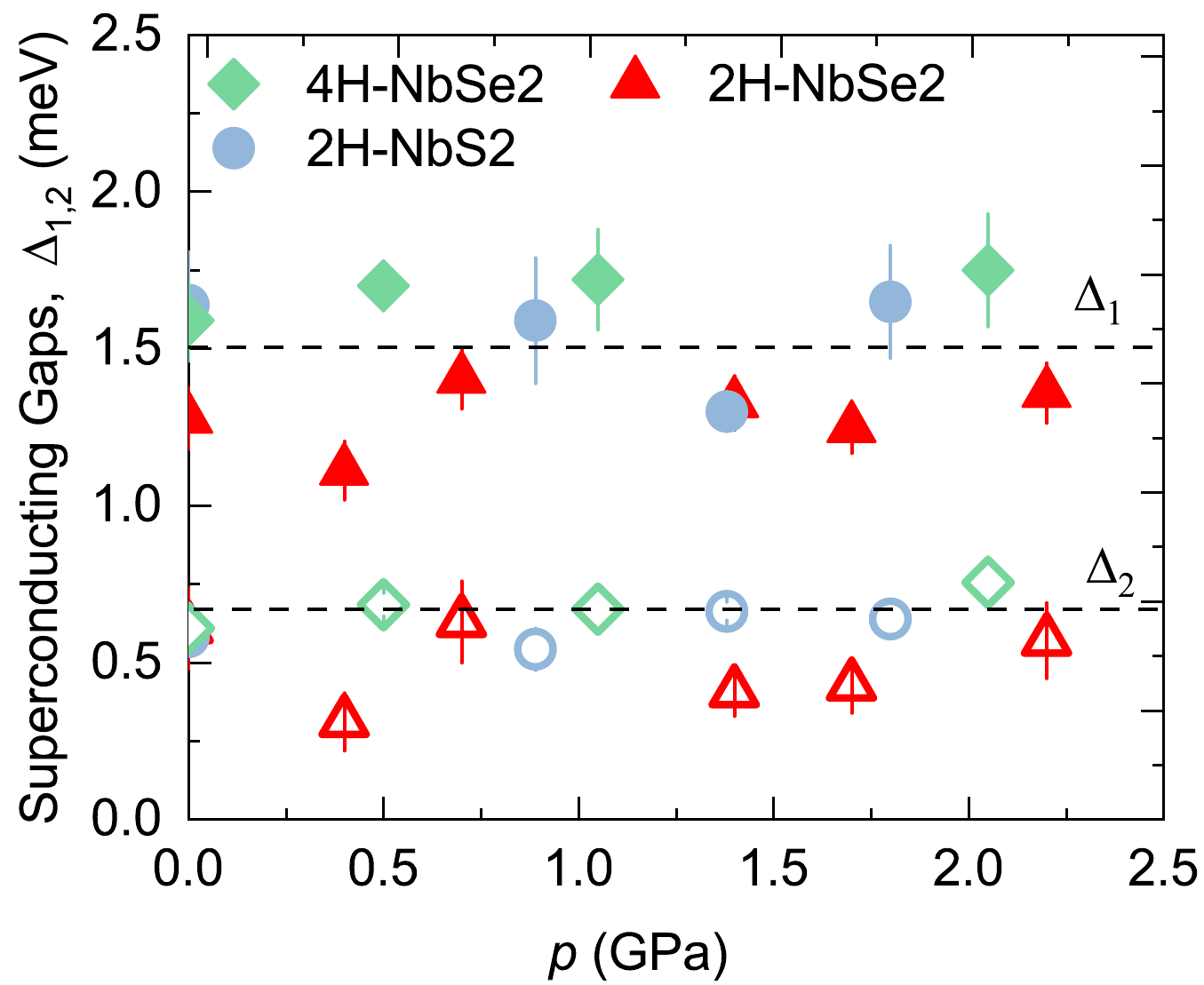}
	\vspace{0cm}
	\caption{\textbf{}Pressure dependence of small superconducting gap ${\Delta}_1$ and the large superconducting gap ${\Delta}_2$. 
	}
	\label{FigS3}
\end{figure*}


\end{document}